# Using I4.0 digital twins in agriculture*

Rodrigo Falcão[1][0000−0003−1222−0046], Raghad Matar[1][0000−0002−7814−1079], and
Bernd Rauch[1][0000−0002−3416−9299]

Fraunhofer IESE, Kaiserslautern, Germany
{rodrigo.falcao, raghad.matar, bernd.rauch}@iese.fraunhofer.de

**Abstract.** Agriculture is a huge domain where an enormous landscape of systems interact to support agricultural processes, which are becoming increasingly digital. From the perspective of agricultural service providers, a prominent challenge is interoperability. In the Fraunhofer lighthouse project Cognitive Agriculture (COGNAC), we investigated how the usage of Industry 4.0 digital twins (I4.0 DTs) can help overcome this challenge. This paper contributes architecture drivers and a solution concept using I4.0 DTs in the agricultural domain. Furthermore, we discuss the opportunities and limitations offered by I4.0 DTs for the agricultural domain.

## 1 Introduction

Digital transformation, i.e., the conception of innovative and digital business models, is taking place and disrupting many major sectors of the economy. We have observed several traditional industries such as transportation, banking, hotel business, and entertainment, that have been impacted by the rise of software companies that approach not simply their products or their processes, but their fundamental businesses digitally [19]. Agriculture, as a major sector of the economy, is already a software-intensive industry where several players cooperate in huge and complex value chains. At the technical level, the digital transformation in agriculture requires digitally available data from the environment, farms, machines, and processes to enable software-supported products and services to work smoothly [25].

However, regarding digital systems, the agricultural domain is rather fragmented: There are various systems, with various data formats, complying with various different standards within its multiple subdomains. Therefore, enabling interoperability in agriculture is challenging. Data is typically distributed in exclusive data storage of suppliers' digital ecosystems. On top of that, there is no or only little semantic interoperability – meaning the ability of applications to exchange data with a shared meaning –, which leads to huge efforts in communication and orchestration for delivering complex end-to-end solutions for farmers [5].

Nowadays, farmers are dealing with several systems to accomplish their goals in all production steps across seasons. Mostly, such systems belong to their own digital

---

ecosystems: Machine manufacturers, for example, usually offer cloud-based solutions to channel data collected from the machinery of their respective fleet, which results in distributed data storages if farmers own machines from different manufacturers. In other cases, systems only cover data from their respective business processes and offer no data exchange across farm systems. As a consequence, farmers often face a situation where they have to use different systems with exclusive data vaults and no or only limited connectivity between those systems. From another perspective, service providers of digital services need to integrate different data sources into their respective service environments in order to offer innovative services for farmers. Therefore, not only farmers but all stakeholders still yearn for a frictionless, yet secure, experience.

In recent years, the idea of digital twins (DTs) in agriculture has been explored (e.g., [22], [25]). The concept of DTs was coined by Michael Grieves in 2003, referring to virtual representations of physical products with two-way communication between them [10]. Initially introduced in the context of product lifecycle management, the concept has evolved, and nowadays DTs refer to the digital representation of any real entity, be it physical (e.g., machines) or not (e.g., services or processes) [15] [16]. However, when it comes to their usage in the agricultural domain, existing research on DTs has not put emphasis on how or whether they could be used to address the interoperability challenge in the domain. DTs have been used to tackle interoperability challenges in another domain, though: In Industry 4.0 (I4.0), the notion of DTs has been realized through one of its core components: the *asset administration shell* (AAS). The term was coined in 2015 [21] in the context of the German research project Platform Industry 4.0[1]. Although the term "digital twin" was not used in the project, over time the convergence of the terms "asset administration shell" and "digital twin" has become evident [26]. In fact, AASs have materialized DTs in the context of I4.0, as declared in [18]: "The Asset Administration Shell helps implementing digital twins for I4.0 and creating interoperability across the solutions of different suppliers.".

Since AASs have been used to enable interoperability in I4.0, in our research we investigated to which extent the same concept could be applied to the agricultural domain, focusing on the quality of interoperability. In this paper, we present architecture drivers for interoperability in the agriculture domain, present a solution concept based on I4.0 DTs, and discuss the opportunities and limitations we found during our investigation.

The remainder of this paper is structured as follows: Section 2 introduces general concepts of the COGNAC project and I4.0 DTs; Section 3 presents related work; Section 4 presents the architecture drivers and a solution concept for I4.0 DTs in the agricultural domain; in Section 5 we discuss our solution; and Section 6 concludes the paper.

## 2    Background

In the Fraunhofer lighthouse project "Cognitive Agriculture" (COGNAC)[2], eight Fraunhofer Institutes have conducted joint research in the area of Smart Farming since 2018.

---
[1] https://www.plattform-i40.de
[2] https://cognitive-agriculture.de

Exploring applied solutions in field automation, novel sensing, smart services, and digital data spaces, the project's core focus has been on the digital transformation of farming processes in the context of evolving digital ecosystems. In this context, different systems, services, and actors interact and collaborate in agricultural processes, building a common *agricultural data space (ADS)*. By analyzing the requirements, we identified interoperability as one of the major challenges in the domain. In our research of suitable solution concepts, we found DTs as a potential approach for coping with interoperability in a digitalized farming setting. In recent years, much research has been conducted on the utilization of DTs for agricultural assets (see Section 3); however, the challenges emerging from interoperability in smart farming have not been covered explicitly. In COGNAC, we drew inspiration from I4.0 and are exploring the use of DTs to realize interoperability between digital services and systems.

The usage of DTs in industry automation has been observed prominently in the context of I4.0 [3]. One example is RAMI 4.0 – the Reference Architectural Model Industry 4.0 [11]. RAMI 4.0 prescribes a layered architecture framework that simultaneously organizes the vocabulary of I4.0 and breaks down the domain complexity into smaller pieces. At the lowest level of the reference architecture, the layer "asset" corresponds to the physical entities (such as machines); right above it, the layer "integration" implements the glue to support the next layer, "communication", which provides access to information. Further layers are "information', "functional", and "business".

DTs are a perfect fit for the realization of the integration layer. Reference implementations are already available. One example is the research project BaSys 4.0[3], which defines a middleware for production systems in the context of I4.0 and puts a strong focus on the implementation of DTs for the production pipeline, where interoperability is an essential quality attribute for high automation levels in processes that take place in a heterogeneous environment. For example, consider a certain production process where different devices cooperate to manufacture lots of a certain item. These devices may come from different producers, use different data formats, and understand different communication protocols. BaSys 4.0 prescribes the implementation of standardized DTs for these devices – the *asset administration shells* –, allowing them to be orchestrated on a higher level of abstraction to enable flexible production lines. This could lead to "lot size one" production being virtually as cost-effective as massive production.

Figure 1 provides a high-level functional view on the usage of I4.0 DTs in a fictional factory, from the factory's perspective. At the bottom layer (Machinery), the physical machines that belong to the factory and are located on the shop floor are represented as computing nodes – in this case, the factory has three machines. In the middle layer (Digital Twins), asset administration shell components implemented in the factory realize digital representations of the machines. Their properties and functions are expressed in terms of *submodels*, as I4.0 calls them. These submodels may or may not be provided by the machine manufacturer – in the former case, it would be good to use them to promote standardization and streamline interoperability; in the latter case, the factory models them as desired. Finally, at the top layer (Orchestration), the DTs are orchestrated into *recipes*, which describe how the different machines should cooperate to implement certain industrial processes.

---

[3] http://basys40.de

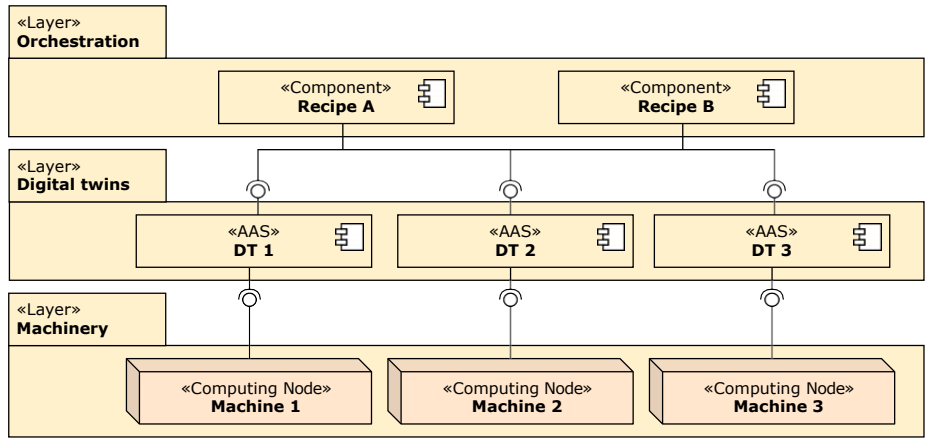

**Fig. 1.** Functional view of an illustrative implementation of DTs in a factory.

This model supports interoperability in different ways. First, asset administration shells have a standardized data structure, so at the technical level, all DTs implemented through asset administration shells are interoperable. Moreover, they work as adapters to convert vendor-specific APIs into a standardized representation of information, which enables syntactic interoperability. Finally, semantic interoperability is supported through references to external type definitions, which should be accessible in order to become part of the domain's shared vocabulary. Such an architecture makes it easy to modify industrial processes by creating new recipes or adjusting existing ones at the orchestration level, despite the potential technical diversity in the factory's machine park, because orchestration happens through the asset administration shells. Furthermore, machines can be replaced at the machinery layer with no impact on the orchestration layer, whereas only minor impact is expected in the digital twin layer if the new machines provide submodels that comply with an established standard for the functionality they offer. In this example, DTs are created to represent devices (i.e., machines in a factory). From a conceptual perspective, however, there could also be DTs for products, processes, or services (systems).

## 3   Related work

To analyze the degree of DT adoption in agriculture, Pylianidis et al. [22] conducted a literature review investigating scientific and gray literature studies published between 2017 and 2020. The results show that the majority of the identified DTs represent agricultural fields, farms, landscapes, and buildings, and are mostly concerned with monitoring and optimization operations. As for the maturity level, most of the identified DTs are only concepts or prototypes. Moreover, the authors conducted a survey case study on the usage of DTs in other disciplines and identified 68 use cases as a result. The results show a delay in the exploration of DTs in agriculture compared to other domains and suggest that the exploration of DTs in agriculture is relatively limited compared with other disciplines. Furthermore, the results reveal that DTs in other disciples provide a wider range of benefits and services. The authors argue that the superficial description of DT applications in the literature has hindered the realization of their benefits and thus slowed their adoption in agriculture.

In another study, Verdouw et al. [25] analyzed how leveraging DTs for farm management can improve productivity and increase sustainability in smart farming. The authors propose a conceptual framework for designing and implementing DTs in smart farming for farm management activities, including planning, monitoring, controlling, and optimizing farm operations. The introduced framework supports the whole lifecycle of physical farm objects and the implementation of the main characteristics of the six DT types in the proposed typology. The authors validated the proposed framework using five smart farming use cases.

Chaux et al. [6] suggest a DT architecture for optimizing the productivity of Controlled Environment Agriculture (CEA) systems by achieving better yield and higher quality of crops with fewer resources. The proposed framework utilizes two simulation softwares as DTs and uses them as test beds for assessing crop treatment and climate control strategies. Furthermore, it has an intelligence layer that generates multiple alternative strategies and selects the optimal one. The authors validated the suggested architecture by building a prototype of an automated greenhouse and performing latency tests to certify the success of the bidirectional communication of the DT architecture.

Moshrefzadeh et al. [17] introduced the notion of Distributed DTs of agricultural landscapes to optimize data integration and thus support diverse stakeholders in establishing a common understanding of the landscape and its objects, and to achieve coordinated decision-making. The concept is part of the Smart Rural Area Data Infrastructure – SRADI, a multidisciplinary information infrastructure developed to handle multiple stakeholders, applications, and distributed information resources [8]. Its core component is the *Catalog*, a metadata registry for distributed landscape resources, such as projects, software, and raw data from landscape objects. The catalog establishes semantic relations between the distributed pieces of information. To achieve interoperability, communication and data modeling are based on open standards. To demonstrate the concept, the authors developed a data infrastructure for the Agricultural Research Center of the Technical University of Munich to support its 30 chair members and organize their cooperation when conducting research on the same land parcels. However, in their proposed solution, the authors achieve interoperability on the level of the metadata and not the data itself.

## 4   Using I4.0 DTs in the agricultural domain

We used the GQM template [4] to frame the goal of our research as follows: "Analyze Industry 4.0 Digital Twins for the purpose of evaluation with respect to interoperability from the point of the view of agricultural service providers in the context of the project COGNAC". To achieve the goal, we had to know the architecture drivers – i.e., the architecture-relevant requirements – related to interoperability in the agricultural domain. Then we explored the solution space, but constrained ourselves to the use of I4.0

DTs, as our goal is to check their adequacy for addressing the interoperability challenge in the agriculture domain.

For eliciting the architecture drivers and designing solutions, we used the software architecture evaluation approach of Knodel and Naab [14]. Each architecture driver was defined in terms of quantified environments, stimuli, and responses, whereas for each design decisions were characterized with opportunities, assumptions, cons/risks, and trade-offs. We assessed drivers and solutions through reviews.

### 4.1   Architecture drivers

Architecture drivers are a particular type of requirements that focus on what matters most for architecture purposes: business goals, constraints, key functional requirements, and quality requirements [14]. Regarding quality requirements, architecture drivers can be expressed as *architecture scenarios*. As defined by Rozanski and Woods, an architecture scenario is "a crisp, concise description of a situation that the system is likely to face, along with a definition of the response required of the system" [23]. Architecture scenarios, also referred to as "quality attribute scenarios" [2], ensure that the quality requirement is expressed in a concrete and measurable way.

From our experience with COGNAC and a feasibility study about data management and Farm Management Information Systems (FMISs) [12], we derived two typical interoperability scenarios in agriculture. In this domain, interoperability relates to software-based systems exchanging data in order to perform certain agricultural processes (e.g., fertilization, weed control, etc.), which are usually implemented by service providers.

**The first scenario (AD.IOP.1) takes place within the boundaries of service providers**. From the point of view of the farmer, who is the end user, they use only one service, provided by one service provider, to perform a certain agricultural process of interest. On their part, service providers often use several systems to implement such agricultural processes, but these systems are not necessarily developed or operated by the service providers, but also by other companies. For example, a service provider who harvests fields may own and use machinery (and corresponding services) built by different manufacturers. Table 1 summarizes AD.IOP.1. Consider, for example, a service provider that offers a weed control service. For the sake of simplification, let's assume that internally, the service provider uses only two systems to implement the weed control process: the "main" service ($Sys_1$), which plans the work and sends instructions to the second system, and a field robot ($Sys_2$), which performs the actual field work and sends the data back to the main service, which in turn generates the work record for the farmer. Consider now that the service provider wants to expand their machine park by acquiring more field robots, but this time from another manufacturer ($Sys_3$). They will therefore have two models of field robots in their machine park, built by different manufacturers, and may use one or another model to execute a farm job, depending on their availability. $Sys_1$ should be able to exchange data with $Sys_3$ (the new field robot) without the need of any design-time change in the implementation of the process.

**The second scenario refers to cross-company data exchange.** In this scenario, the farmer is aware of the fact that they are using more than one system and explicitly authorizes the data exchange. Examples are FMISs and fertilization recommendation

Using I4.0 DTs in agriculture    7**Table 1.** AD.IOP.1: Intra-company data exchange.

|  | Description | Quantification |
|---|---|---|
| **Environment** | · Service provider $SP_1$ uses two systems, $Sys_1$ and $Sys_2$, to perform a certain agricultural process $P$, which is controlled via software.<br>· $Sys_1$ and $Sys_2$ are able to exchange data. | n/a |
| **Stimulus** | · The service provider replaces $Sys_2$ with $Sys_3$, which is maintained by a third-party company. | n/a |
| **Response** | · $Sys_1$ and $Sys_3$ are able to exchange data | · No modification at design time is required for the implementation of the process $P$. |

**Table 2.** AD.IOP.2: Cross-company data exchange.

|  | Description | Quantification |
|---|---|---|
| **Environment** | · Service $S_1$ needs field data to provide its service. | · $S_1$ needs one-time access to field data |
| **Stimulus** | · $S_1$ requests field data from the new FMIS | n/a |
| **Response** | · $S_1$ receives and can understand the field data from the new FMIS | · No modification at design time is required in $S_1$ to be able to interoperate with the new FMIS. |

services. A farmer $F_1$ may use $FMIS_1$, provided by service provider $SP_1$, and the fertilization recommendation service $FRS_1$ to get recommendations. $FRS_1$ requires field data to provide the recommendation, so it should somehow get the data from $FMIS_1$. Next year, the farmer may decide to change either their FMIS or their fertilization recommendation service (or both). In such a situation, they still want their service providers to be able to exchange data as before. Table 2 summarizes AD.IOP.2. In this scenario, a service provider operates the service $S_1$, which is already established in the market. In order to provide its service, $S_1$ needs one-time access to read certain field data, which in turn is managed by the farmer through their FMIS. $S_1$ is already capable of getting field data from the lead FMISs on the market; however, a new FMIS now enters the market and starts to gain popularity among farmers. One of these early adopters of the new FMIS wants to use $S_1$. Assuming that all accesses have already been granted, $S_1$ should be able to retrieve the required field data from the new FMIS without the need for any design-time modification.

### 4.2 Solution concept

Concerning the intra-company interoperability issue, we envision the *service provider* in agriculture as analog to the *factory* in I4.0. In a factory (company), several machines (production means) are orchestrated to perform industrial activities (process), which results in the manufacturing of industrial goods (product). Conversely, in the agricultural domain, service providers (company) use several systems (production means) to perform agriculture-related activities (processes). When it comes to the results, they are not necessarily tangible, though: The process may act directly on the field, which has an impact on its status – for example, after a certain agriculture process, a field may be fertilized, protected, or harvested, among other possibilities. It would also be possible to think of the crop or the yield as the product that is impacted by the agricultural process, even though such impact is not necessarily perceived immediately (e.g., a fertilized field will influence the growth of the plants, but this can only be observed afterwards). On the other hand, there are agricultural process that do not produce tangible (i.e., physical) results at all, for example recommendation processes, which are quite common in the agricultural domain.

**The first design decision (DD.1) is** *the usage of AASs as representatives of all software systems that are to interoperate in the ADS*. This decision can be traced directly to the architecture solution blueprint for I4.0 applications introduced by Antonino et al. [1], who make a case for the "use of Digital Twins as digital representatives of the different physical and logical entities involved in the production process". In our case, we are talking about *logical entities*: the systems. Each AAS will therefore work as an Adapter (structural architectural pattern described by Gamma et al. [9]) between the software system and those who want to interoperate with it. Opportunities, assumptions, risks, and trade-offs are presented in Table 3.

Table 3. DD1: Implementation of AASs as representatives of all systems in the ADS.

| | |
|---|---|
| **Opportunities/pros** | AASs provide standardized interfaces and data structure constructs, so from a technical point of view, all systems will expose their properties and functionalities in the same way, making it easier for clients to communicate with them. |
| **Assumptions** | The constructs available to create the submodels (and therefore to design the AASs) are generic enough to express the variety of properties and functions in the agricultural domain. |
| **Cons/Risks** | The usage of generic constructs to describe the submodels might result in sub-optimal data structures |
| **Trade-offs** | The overhead required at design time to implement the *digital twin layer* (see Figure 2). Furthermore, some impact may be noted at runtime from the required data transformations. |

Figure 2 illustrates a functional view of the implementation of DTs in an automated weed control company specializing in weed control for both potato and sugar beet crops. Consider that the service provider has a main Weed Control Operation System, which is used by their employees to plan and carry out the jobs. Let's assume that internally, the company uses two additional systems (production means) to support their work: field robots, which are programmed to do the actual job on the field and collect data about it, and a route planner, which calculates the optimal route for the field robots. In the orchestration layer, the service provider can program two *recipes* that implement the weed control processes; i.e., they describe how to perform weed control on a potato crop and on a sugar beet crop. These recipes would rely on DTs of the *production means*: the route planner system and the field robot. In the example, two field robots are available. Their respective DTs should implement the same interface (see DTs layer) that is used in the orchestration layer. Finally, in the systems layer, each individual system exposes its specific interfaces, which are used directly in the implementation details of their respective DTs.

**Table 4.** DD2: Implementation of AASs as representatives of fields.

| | |
|---|---|
| Opportunities/pros | Digital field twins represent a single source of truth for field-related data, making data available in a standardized, up-to-date, and non-redundant way. All services can get field data from one place. |
| Assumptions | The digital fields twins are hosted in a infrastructure that is reachable through the Internet, and both hosts and digital field twins are uniquely identified. |
| Cons/Risks | In case a digital field twin is unavailable, all inter-company interoperability that depends on the corresponding field data is affected. |
| Trade-offs | Additional (dedicated) infrastructure, along with the associated costs for implementation, operation, and maintenance, is required to deploy digital field twins. |

**Table 5.** DD3: Use of a mediator to process field-related data exchange requests.

| | |
|---|---|
| Opportunities/pros | Systems that are interested in field data do not have to comply with a given specific interface provided by the digital field twin. New field data can be incorporated into the twin through generic commands. |
| Assumptions | All participants agree on the usage of an open shared vocabulary, providing the data with semantics. |
| Cons/Risks | There is a risk that the usage of generic constructs will make the usage of the mediator complex, depending on the data involved. |
| Trade-offs | The complexity of the interaction between systems is replaced by the complexity of the mediator itself (as foreseen in [9]). Furthermore, the overhead caused by data transformations and the need for multiple calls to transfer complex data can impact performance. |

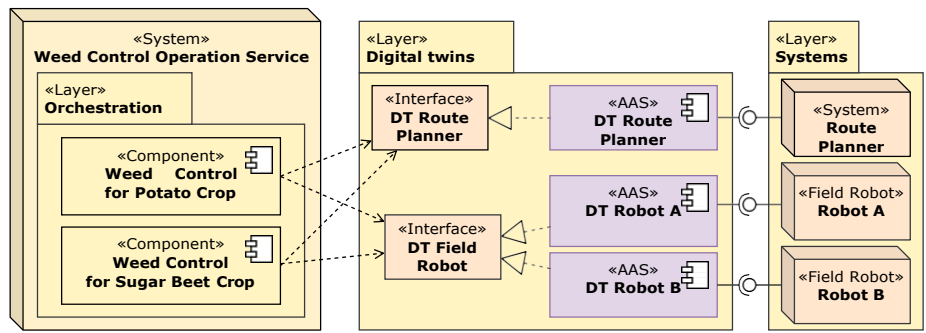

**Fig. 2.** Functional view of the usage of DT by a weed control company.

The service provider is in the best position to define how the data must flow among participants in the ecosystem, so they should implement the recipes for performing different types of jobs (one recipe for each crop type). If there is then a business need to change the process (maybe to allow working on a third crop type), a new recipe is created at the orchestration level. Conversely, if there is a need to replace Robot A with Robot B, the recipe remains untouched, since both robots, potentially from different manufacturers, are accessed through the standardized DT layer.

When it comes to inter-company interoperability, our experience not only in the project COGNAC but also in multiple industry projects has revealed the prominent role of field-related data in the agricultural domain. Service providers need field data, which is typically stored in more than one FMIS used by farmers, in order to provide their services. Field data has different dimensions, including geographic (e.g., field boundaries and terrain slope), environmental (e.g. weather – past, current, and forecast), agronomic (e.g., soil nutrient levels and plant health status). Having digital representations of the systems involved would not suffice to enable adequate interoperability regarding field-related data because the data is scattered across several FMISs, which jeopardizes data qualities such as completeness, currency, consistency, and availability [13]. This led us to **the second design decision (DD.2)**: *the use of AASs to implement digital representatives of fields* – referred to as *digital field twins*. When providing a service, the service provider should acquire the needed field data from the digital field twin; if the corresponding services generate field-related data, the data should also be sent to and stored in the digital field twin. To realize this, we introduce the concept of TwinHub, a vendor-neutral digital platform that hosts a farmer's digital field twins. Table 4 characterizes this design decision.

The implementation of digital field twins, as described to this point, has an additional assumption that has not been made explicit in Table 4. Systems that need field data can benefit from the digital field twin only if they know the interface of the digital field twin in advance , whereas such an interface should statically reflect how digital field twins expose a representation of its data model. However, this may or may not be a valid assumption. If the client knows the interface provided by the digital field twin and complies with it, syntactic interoperability – where two systems can exchange data because there is a known data structure [24] – can be achieved. The digital field twin could be accessed from a recipe in the orchestration layer of a certain system and the field data would be available to support the process described in the corresponding recipe.

Conversely, if the data structure of the digital field twin cannot be known (or even defined) at design time, we need something else to enable interoperability, which takes us to **our third design decision (DD.3)**: *the use of a mediator to process data exchange requests between digital field twins and other digital twins*. The mediator is a self-contained service that must know the reflexive syntax of AASs in order to be able to call any AAS. This solution combines characteristics of the architectural patterns *Mediator* and *Command* [9]: It centralizes the control of communication between several parties and exposes an API to receive data exchange requests (e.g., "get *field boundaries* and *crop type* from *digital field twin A* and send it to *digital twin B*"). Such an interface should be called by the system that wants to exchange data. Table 5 summarizes the characteristics of this decision, and Figure 3 illustrates both DD.2 and DD.3. All participant systems use a reference vocabulary and request field-related data exchanges through the Mediator, which in turn accesses the digital field twin and the corresponding DTs of each participant system.

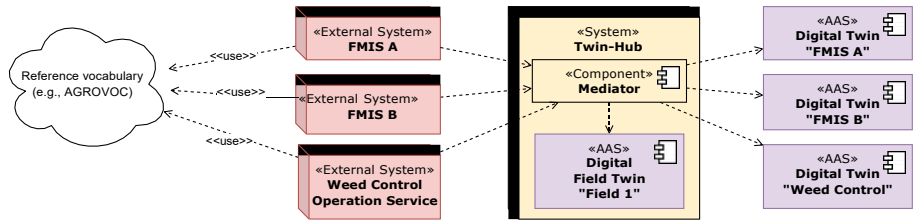

**Fig. 3.** Functional view of the use of digital field twins (DD.2) and a mediator (DD.3).

## 5   Discussion

We have investigated the usage of I4.0 DTs in agriculture, which may raise the question of whether DTs are a good idea for this domain in the first place. The increasing interest in the topic that can be observed in recent years has indicated so. DTs have generally been perceived as an enabler for interoperability [20]. In the agricultural domain, their importance has grown as well, although it has not yet reached maturity. There is still disagreement on essential aspects, such as to which extent real-time data synchronization between DTs and real entities is needed [25] or not [22]. Moreover, with respect to technical aspects of the realization of DTs in agriculture, the results are usually presented at the conceptual level through illustrative use cases; i.e., they do not explain how interoperability can be achieved technically. In the current state of digital farming, the few implementations of DTs are limited to single-system contexts and there is no overall concept for collaboratively utilizing DTs across system boundaries.

This is where, from our point of view, I4.0 DTs can offer their maturity to support the agricultural domain. In I4.0, stakeholders have agreed upon a reference model and have developed not only reference architectures but also reference implementations for DTs (e.g., [7]). Following the idea of having DTs to represent the production means, we looked at systems that contribute to agricultural processes as production means, which led us to decide to create DTs for these systems. Since DTs are usually thought of as digital representatives of *physical* entities, we considered the creation of *DTs for systems* counter-intuitive at first though. On the other hand, the idea of having *digital field twins* was mostly straightforward. The field is a central entity in the context of a farm and field data plays a central role in all agricultural processes. Furthermore, fields are physical entities. Still, we should remember that the DT definition comprises a closed loop between the real and the digital entity (i.e., changes in the real entity are reflected in the DT, and vice-versa). It seems to be clear that when the value of a soil nutrient such as nitrogen changes in the real field, this information can be captured by sensors and reflected in the digital field twin. However, how could it be possible to increase, for instance, the nitrogen level in the digital field twin and have it reflected in the real entity? One solution could be to have the digital field twin trigger an agricultural process (e.g., fertilization) that would, in turn, raise the nitrogen level.

Apart from technical aspects, one of the main drivers of DT development in I4.0 is the market players' demand for interoperability and standardization. In agriculture, however, the digital ecosystem is still lacking a common interest in comprehensive interoperability as some market players are reluctant to collaborate with others and try to gain market shares within their own digital ecosystems. The implementation of the proposed utilization of I4.0 DTs would require domain-wide agreement on a common technological framework, which poses a major challenge for its success. The technical solution approach could initially be implemented by smaller parts of the domain and be expanded later. There are also open organizational questions, such as who would implement and operate data hubs offering digital field twin interfaces (such as the TwinHub).

Among the limitations we identified during our investigation is the need for better support for cross-twin operations. I4.0 seems to work on the assumption that all DTs that are needed to support a certain process are known and can therefore be discovered directly and integrated into a recipe. However, for digital field twins it seems to be necessary to first search for and filter sets of twins. As far as we know, such higher-level operations are not yet supported. Another limitation is that it is not possible to access properties via their semantic definitions (even though it is possible to add semantic annotations to the properties).

*Threats to validity.* Threats to validity of this work include the usage of reviews, which are limited methods [14], to assess drivers and solutions. It is also worth noting that our analysis may be biased by the characteristics of Eclipse BaSyx 1.0 [7], a specific AAS implementation, which we used in the early development stages of our prototypes.

## 6   Conclusion

In this paper, we described two architecture drivers for interoperability in the agricultural domain and presented a solution concept using I4.0 DTs. To the best of our knowledge, no previous study in the literature has investigated the usage of I4.0 DTs for achieving interoperability on the data exchange level among the different actors in the agricultural domain. From a technical point of view, we believe that smart farming can move towards concrete solutions by benefiting from the progress already made by I4.0 DTs.

As future work, we will further develop our current prototype to increase the level of confidence in the solution, focusing on digital field twins, which have a more obvious impact on inter-company interoperability.

## References

1. Antonino, P.O., Schnicke, F., Zhang, Z., Kuhn, T.: Blueprints for architecture drivers and architecture solutions for industry 4.0 shopfloor applications. In: ECSA 2019 - Volume 2. pp. 261–268 (2019)
2. Bass, L., Clements, P., Kazman, R.: Software architecture in practice. Addison-Wesley Professional (2003)
3. Bauer, T., Antonino, P.O., Kuhn, T.: Towards architecting digital twin-pervaded systems. In: SESoS 2019 and WDES 2019. pp. 66–69. IEEE (2019)
4. Caldiera, V.R.B.G., Rombach, H.D.: The goal question metric approach. Encyclopedia of software engineering pp. 528–532 (1994)
5. Calvet, E., Falcão, R., Thom, L.H.: Business process model for interoperability improvement in the agricultural domain using digital twins. In: PACIS 2022. AIS (2022)
6. Chaux, J.D., Sanchez-Londono, D., Barbieri, G.: A digital twin architecture to optimize productivity within controlled environment agriculture. Applied Sciences **11**(19), 8875 (2021)
7. Eclipse: Eclipse BaSyx. https://projects.eclipse.org/projects/dt.basyx (2022), accessed: 2022-07-06
8. Gackstetter, D., Moshrefzadeh, M., Machl, T., Kolbe, T.H.: Smart rural areas data infrastructure (sradi)–an information logistics framework for digital agriculture based on open standards. 41. GIL-Jahrestagung, Informations-und Kommunikationstechnologie in kritischen Zeiten (2021)
9. Gamma, E., Helm, R., Johnson, R., Johnson, R.E., Vlissides, J., et al.: Design patterns: elements of reusable object-oriented software. Pearson Deutschland GmbH (1995)
10. Grieves, M.: Digital twin: manufacturing excellence through virtual factory replication. White paper **1**, 1–7 (2014)
11. Hankel, M., Rexroth, B.: The reference architectural model industrie 4.0 (rami 4.0). ZVEI **2**(2), 4–9 (2015)
12. Herlitzius, T., Schroers, J.O., Seuring, L., Striller, B., Henningsen, J., Jeswein, T., Neuschwander, P., Rauch, B., Scherr, S.A., Martini, D., et al.: Betriebliches datenmanagement und fmis (2022)
13. ISO/IEC 25024:2015. Systems and software engineering – Systems and software Quality Requirements and Evaluation (SQuaRE) – Measurement of data quality. Standard, ISO (2015)
14. Knodel, J., Naab, M.: Pragmatic Evaluation of Software Architectures, vol. 1. Springer (2016)
15. Kuhn, T.: Digitaler Zwilling. Informatik-Spektrum **40**(5), 440–444 (2017)
16. Malakuti, S., van Schalkwyk, P., Boss, B., Sastry, C., Runkana, V., Lin, S., Rix, S., Green, G., Baechle, K., Varan Nath, S.: Digital twins for industrial applications: Definition. Business Values, Design Aspects, Standards and Use Cases: An Industrial Internet Consortium Whitepaper (2020)
17. Moshrefzadeh, M., Machl, T., Gackstetter, D., Donaubauer, A., Kolbe, T.H.: Towards a distributed digital twin of the agricultural landscape. JoDLA **5**, 173–118 (2020)
18. Neidig, J., Orzelski, A., Pollmeier, S.: Asset Administration Shell – Reading Guide. https://www.plattform-i40.de/IP/Redaktion/DE/Downloads/Publikation/AAS-ReadingGuide_202201.html (2022), accessed: 2022-06-02
19. Parker, G.G., Van Alstyne, M.W., Choudary, S.P.: Platform revolution: How networked markets are transforming the economy and how to make them work for you. WW Norton & Company (2016)
20. Piroumian, V.: Digital twins: Universal interoperability for the digital age. Computer **54**(1), 61–69 (2021)

14      R. Falcão et al.

21. Platform Industrie 4.0: Umsetzungsstrategie Industrie 4.0: Ergebnisbericht der Plattform Industrie 4.0. https://www.plattform-i40.de/IP/Redaktion/DE/Downloads/Publikation/umsetzungsstrategie-2015.html (2015), accessed: 2022-06-02
22. Pylianidis, C., Osinga, S., Athanasiadis, I.N.: Introducing digital twins to agriculture. Computers and Electronics in Agriculture **184**, 105942 (2021)
23. Rozanski, N., Woods, E.: Software systems architecture: working with stakeholders using viewpoints and perspectives. Addison-Wesley (2012)
24. Valle, P.H.D., Garcés, L., Nakagawa, E.Y.: A typology of architectural strategies for interoperability. In: SBCARS 2019. pp. 3–12 (2019)
25. Verdouw, C., Tekinerdogan, B., Beulens, A., Wolfert, S.: Digital twins in smart farming. Agricultural Systems **189**, 103046 (2021)
26. Wagner, C., Grothoff, J., Epple, U., Drath, R., Malakuti, S., Grüner, S., Hoffmeister, M., Zimermann, P.: The role of the industry 4.0 asset administration shell and the digital twin during the life cycle of a plant. In: ETFA 2017. pp. 1–8. IEEE (2017)